\documentclass[conference]{IEEEtran}
\usepackage[utf8]{inputenc}
\usepackage[frozencache]{minted}
\usepackage{url}
\usepackage{xspace}
\usepackage[svgnames]{xcolor}
\usepackage{graphicx}
\usepackage{booktabs}
\usepackage{sparklines}

\usepackage{enumitem}
\usepackage{tikz}
\usetikzlibrary{calc,positioning}

\usepackage[para,online,flushleft]{threeparttable}
\usepackage{tabularx}
\usepackage{tabulary}
\usepackage{colortbl}

\usepackage[capitalize]{cleveref}

\makeatletter
\let\MYcaption\@makecaption
\makeatother

\usepackage[font=footnotesize]{subcaption}

\makeatletter
\let\@makecaption\MYcaption
\makeatother

\usepackage{zi4}

\newcommand{\tool}{Transfer Tutor}
\newcommand{\expert}{Julie}

\IEEEoverridecommandlockouts
\IEEEpubid{\makebox[\columnwidth]{978-1-5386-4235-1/18/\$31.00
\copyright2018
IEEE \hfill} \hspace{\columnsep}\makebox[\columnwidth]{}}

\begin{document}

\title{It's Like Python But: Towards Supporting Transfer of Programming Language Knowledge}

\author{\IEEEauthorblockN{Nischal Shrestha}
\IEEEauthorblockA{NC State University\\
Raleigh, NC, USA\\
nshrest@ncsu.edu}
\and
\IEEEauthorblockN{Titus Barik}
\IEEEauthorblockA{Microsoft\\
Redmond, WA, USA\\
titus.barik@microsoft.com}
\and
\IEEEauthorblockN{Chris Parnin}
\IEEEauthorblockA{NC State University\\
Raleigh, NC, USA\\
cjparnin@ncsu.edu}
}

\maketitle

% TODO(tbarik): Inspiration
% https://www.rubysteps.com/why-learning-your-second-programming-language-can-be-harder-than-the-first/

\begin{abstract}
% Abstract

% Last sentence not very clear
% First paragraph is good

Expertise in programming traditionally assumes a binary novice-expert divide. Learning resources typically target programmers who are learning programming for the first time, or expert programmers for that language. An underrepresented, yet important group of programmers are those that are experienced in one programming language, but desire to author code in a different language. For this scenario, we postulate that an effective form of feedback is presented as a transfer from concepts in the first language to the second. Current programming environments do not support this form of feedback.

In this study, we apply the theory of learning transfer to teach a language that programmers are less familiar with--–such as R--–in terms of a programming language they already know--–such as Python. We investigate learning transfer using a new tool called \tool{} that presents explanations for R code in terms of the equivalent Python code. Our study found that participants leveraged learning transfer as a cognitive strategy, even when unprompted. Participants found \tool{} to be useful across a number of affordances like stepping through and highlighting facts that may have been missed or misunderstood. However, participants were reluctant to accept facts without code execution or sometimes had difficulty reading explanations that are verbose or complex. These results provide guidance for future designs and research directions that can support learning transfer when learning new programming languages.

% Finally, a qualitative evaluation identified features such as highlighting syntax elements, explanations that related unfamiliar R concepts to Python, and stepping through the code, as being useful to programmers.
\end{abstract}

\section{Introduction}

Programmers are expected to be fluent in multiple programming languages. When a programmer switches to a new project or job, there is a ramp-up problem where they need to become proficient in a new language~\cite{sim1998ramp}. For example, if a programmer was proficient in Python, but needed to learn R, they would need to consult numerous learning resources such as documentation, code examples, and training lessons. Unfortunately, current learning resources typically do not take advantage of a programmer's existing knowledge and instead present material as if they were a novice programmer~\cite{loksa2016programming}. This style of presentation does not support experienced programmers~\cite{berlin1993} who are already proficient in one or more languages and harms their ability to learn effectively and efficiently~\cite{kalyuga:expertise}.

Furthermore, the new language may contain many inconsistencies and differences to previous languages which actively inhibit learning. For example, several blogs and books~\cite{burns:inferno} have been written for those who have become frustrated or confused with the R programming language. In an online document~\cite{arggh}, Smith lists numerous differences of R from other high-level languages which can confuse programmers such as the following:

% \begin{quote}\itshape
% If you are using R and you think you're in hell, this is a map for you. A book about trouble spots, oddities, traps, glitches in R.
% \end{quote}

% \begin{quote}
% \itshape
% Dots in identifier names are just part of the identifier. They are not scope operators. They are not operators at all. They are just a legal character to use in the names of things.
% \end{quote}

\begin{quote}
\itshape
Sequence indexing is base-one. Accessing the zeroth element does not give an error but is never useful.
\end{quote}

In this paper, we explore supporting learning of programming languages through the lens of \emph{learning transfer}, which occurs when learning in one context either enhances (positive transfer) or undermines (negative transfer) a related performance in another context~\cite{Perkins:transfer}. Past research has explored transfer of cognitive skills across programming tasks like comprehension, coding and debugging~\cite{pirolli:transfer, kessler1988transfer, pennington1995transfer}. There has also been research exploring the various difficulties of learning new programming languages \cite{Scholtz:subsequent, quanfeng} and identifying programming misconceptions held by novices~\cite{kaczmarczyk2010identifying}. However, limited research has focused on the difficulties of learning languages for experienced programmers and the interactions and tools necessary to support transfer.

% We built a new training tool called \tool{} designed to facilitate near transfer (to closely related contexts). We use a new technique called ``interactive delta learning" which guides programmers through code snippets of two programming languages and highlights reusable concepts from a familiar language to learn a new language. \tool{} also warns programmers about potential misconceptions. In the long-term, we would like to support the ability to assist programmers transferring knowledge between multiple languages.

% feedback edit
% We built and evaluated a new training tool called \tool{} to support learning transfer of programming knowledge. We propose an approach called \emph{interactive delta learning} which guides programmers through code snippets of two programming languages and highlights reusable concepts from a familiar language to learn a new language. \tool{} also warns programmers about potential misconceptions carried over from the previous language. In the long-term, we would like to support the ability to help programmers rapidly obtain proficiency in new technology by leveraging learning transfer.

To learn how to support transfer, we built a new training tool called \tool{} that guides programmers through code snippets of two programming languages and highlights reusable concepts from a familiar language to learn a new language. \tool{} also warns programmers about potential misconceptions carried over from the previous language~\cite{Fix}.

We conducted a user study of \tool{} with 20 participants from a graduate Computer Science course at North Carolina State University. A qualitative analysis on think-aloud protocols revealed that participants made use of learning transfer even without explicit guidance. According to the responses to a user satisfaction survey, participants found several features useful when learning R, such as making analogies to Python syntax and semantics. However, participants also pointed out that \tool{} lacks code executability and brevity. Despite these limitations, we believe a learning transfer tool can be successful in supporting expert learning of programming languages, as well as other idioms within the same language. We discuss future applications of learning transfer in other software engineering contexts, such as assisting in code translation tasks and generating documentation for programming languages.

% TODO(tbarik): These aren't actually contributions. These are just things you did. Need what we learned (3) is probably okay.
%\begin{itemize}
    %\item The transfer training tool is effective in teaching R.
    %\item Programmers use learning transfer when learning a new programming language.
    %\item The tool supports learning transfer.
        
%\end{itemize}

\section{Motivating Example}
\label{sec:motivating_example}

% TODO(tbarik): This needs to be subfig package, not subcaption due
% to custom IEEE captions.
\begin{figure*}[ht]
    \centering
    \begin{tabular}{c|c}
        \begin{subfigure}[b]{0.45\textwidth}
            \centering
            \begin{minted}[fontsize=\footnotesize,linenos,autogobble,breaklines,breakafter=/]{Python}
                df = pd.read_csv('Questions.csv')
                df = df[df.Score > 0][0:5]
            \end{minted}
        \caption{Python}
        \label{fig:motivation:Python}
        \end{subfigure}
        &
        \begin{subfigure}[b]{0.45\textwidth}
        \centering
        \begin{minipage}{.8\textwidth}
            \begin{minted}[fontsize=\footnotesize,linenos,autogobble]{Dockerfile}
                df <- read.csv('Questions.csv')
                df <- df[df$Score > 0, ][1:5, ]
            \end{minted}
        \end{minipage}
        \caption{R}
        \label{fig:motivation:R}
        \end{subfigure}
\end{tabular}
\caption{(a) Python code for reading data, filtering for positive scores and selecting 5 rows. (b) The equivalent code in R. }
\label{fig:motivation}
\end{figure*}  

% motivating example

% you’re missing the step of importing pandas package
% example should just be example, move last paragraph to next section

Consider Trevor, a Python programmer who needs to switch to R for his new job as a data analyst. Trevor takes an online course on R, but quickly becomes frustrated as the course presents material as if he is a novice programmer and does not make use of his programming experience with Python and Pandas, a data analysis library. Now, Trevor finds himself ill-equipped to get started on his first task at his job, tidying data on popular questions retrieved from Stack Overflow (SO), a question-and-answer (Q\&A) community~\cite{stackoverflow}. Even though he is able to map some concepts over from Python, he experiences difficulty understanding the new syntax due to his Python habits and the inconsistencies of R. Trevor asks help from \expert{}, a seasoned R programmer, by asking her to review his R script (see \cref{fig:motivation}) so he can learn proper R syntax and semantics.

Trevor's task is to conduct a typical data analysis activity, tidying data. He is tasked with the following: 1) read in a comma-separated value (csv) file containing Stack Overflow questions 2) filter the data according to positive scores and 3) select the top five rows. \expert{} walks him through his Python code and explains how they relate to the equivalent code she wrote in R.

\expert{} teaches Trevor that R has several assignment operators that he can use to assign values to variables but tells him that the \texttt{<-} syntax is commonly used by the R community. However, she tells him that the \texttt{=} operator can also be used in R just like Python. To read a csv file, \expert{} instructs Trevor to use a built-in function called \texttt{read.csv()} which is quite similar to Python's \texttt{read\_csv()} function.

Moving on to the next line, \expert{} explains that selecting rows and columns in R is very similar to Python with some subtle differences. The first subtle difference that she points out is that when subsetting (selecting) rows or columns from a data frame in Python, using the \texttt{[} syntax selects rows. However, using the same operator in R will select columns. \expert{} explains that the equivalent effect of selecting rows works if a comma is inserted after the row selection and the right side of the comma is left empty (Figure \ref{fig:motivation:R}). \expert{} tells him that since the right side is for selecting columns, leaving it empty tells R to select all the columns. To reference a column of a data frame in R, \expert{} explains that it works almost the same way as in Python, except the \texttt{.} (dot) must be replaced with a \texttt{\$} instead. Finally, \expert{} points out that R's indexing is 1-based, so the range for selecting the five rows must start with 1, and unlike Python, the end index is inclusive. Trevor now has some basic knowledge of R. Could tools help Trevor in the same way \expert{} was able to?

% replace "the tool" with Transfer Tutor where appropriate
% \section{Transfer Training Tool}
\section{\tool}

\label{sec:thetool}

\subsection{Design Rationale}
We created a new training tool called \tool{} that takes the place of an expert like \expert{} and makes use of learning transfer to teach a new programming language. \tool{} teaches R syntax and semantics in terms of Python to help provide scaffolding~\cite{sawyer2005cambridge} so programmers can start learning from a familiar context and reuse their existing knowledge. Our approach is to illustrate similarities and differences between code snippets in Python and R with the use of highlights on syntax elements and different types of explanations.

We designed \tool{} as an interactive tool to promote ``learnable programming"~\cite{victor2012} so that users can focus on a single syntax element at a time and be able to step through the code snippets on their own pace. We made the following design decisions to teach data frame manipulations in R: 1) highlighting similarities between syntax elements in the two languages 2) explicit tutoring on potential misconceptions and 3) stepping through and highlighting elements incrementally.

% implementation and design decisions 
% We created \tool{} as a research tool to investigate interactive delta learning through the following design decisions, applied to the topic of data frame manipulation: 1) highlighting similarities between syntax elements to support learning transfer 2) explicit tutoring on potential misconceptions 3) stepping through and highlighting elements of the snippets incrementally. \tool{} is an interactive guide that presents a series of lessons that build on each other. An interactive tool is chosen over a traditional document to promote “learnable programming”~\cite{victor2012} to clearly communicate the concepts of the new language and help learners focus on one syntax element at a time. 

% We have designed a tool that attempts to take the place of an expert like Julia and teach R to programmers by making use of transfer. We accomplish this by presenting highlights of positive and negative transfers between two equivalent code snippets written in the two programming languages. 

\subsection{Learning Transfer}

% We applied the theory of learning transfer through these feedback mechanisms in the interface:
\tool{} supports learning transfer through these feedback mechanisms in the interface:

\begin{itemize}
    \item \textbf{Negative Transfer:} `Gotchas' warn programmers about a syntax or concept that either does not work in the new language or carries a different meaning and therefore should be avoided.
    \item \textbf{Positive Transfer:} `Transfer' explanations describe a syntax or concept that maps over to the new language.
    \item \textbf{New Fact:} `New facts' describe a syntax or concept that has little to no mapping to the previous language.
\end{itemize}

% Each type of feedback consists of a highlighted portion of the code in the associated language (Python or R) with its respective explanation, which serves as affordances~\cite{Greeno1993} for learning transfer. Furthermore, we support deliberate connections between elements, by allowing participants to step through the code, which helps them make a mindful abstraction of the concepts~\cite{Schunk}. Finally, we focus on transferring declarative knowledge~\cite{harvey:transfer}, such as syntax rules, rather than procedural knowledge, such as general problem-solving strategies. The next section will present a use case of the tool.

% camera-ready edit
% Each type of feedback consists of a highlighted portion of the code in the associated language (Python or R) with its respective explanation, which serves as affordances~\cite{Greeno1993} for learning transfer. Furthermore, we support deliberate connections between elements, by allowing participants to step through the code, which helps them make a mindful abstraction of the concepts~\cite{Schunk}. Finally, we focus on transferring declarative knowledge~\cite{harvey:transfer}, such as syntax rules, rather than procedural knowledge, such as general problem-solving strategies. The next section will present screenshots and a scenario of using \tool{} to learn R. The figures have been annotated with arrows and text labels to point out the various features of the tool and are not actually presented to the users.

% feedback edit
Each type of feedback consists of a highlighted portion of the code in the associated language (Python or R) with its respective explanation, which serves as affordances for transfer~\cite{Greeno1993}. Furthermore, we support deliberate connections between elements, by allowing participants to step through the code, which helps them make a mindful abstraction of the concepts~\cite{Schunk}. Finally, we focus on transferring declarative knowledge~\cite{harvey:transfer}, such as syntax rules, rather than procedural knowledge, such as general problem-solving strategies.

\subsection{User Experience}

This section presents screenshots of \tool{} and a use case scenario. The user experience of \tool{} is presented from the perspective of Trevor who decides to use the tool to learn how to select columns of a data frame in R, a 2D rectangular data structure which is also used in Python/Pandas. The arrows and text labels are used to annotate the various features of the tool and are not actually presented to the users.

% We tried to ensure that the programmer can easily read both of the code snippets in one area by aligning them vertically and minimizing the distance between the snippets. The top to bottom layout was chosen as opposed to a left to right layout so that the reader would not have to shift their gaze back and forth across the screen, which makes it harder to keep track of the highlighted element on each sidee

\subsubsection{Code Snippets and Highlighting}

Trevor opens up \tool{} and notices that the tool displays two lines of code, where the top line is Python, the language that he is already familiar with and on the bottom is the language to learn which is R. Trevor examines the stepper buttons below the snippets and clicks \tikz\draw node[circle, draw=black, fill=black, text=white, thick, minimum size=0.15pt, font=\tiny,inner sep=1.25pt] (current) {\textbf{3}}; which begins the lesson and highlights some syntax elements:

\begin{center}
\includegraphics[width=\linewidth]{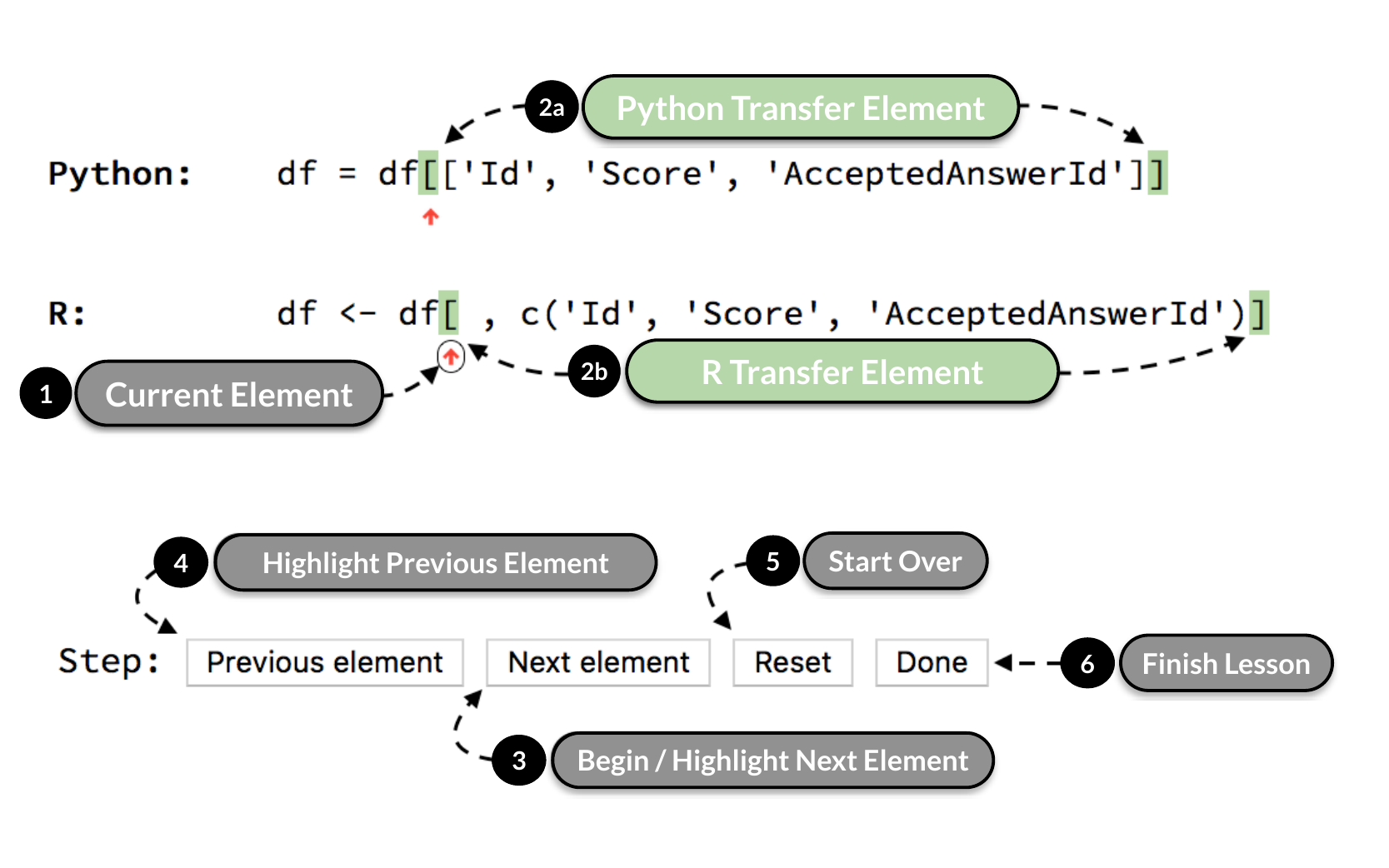}
\end{center}

Travis notices \tikz\draw node[circle, draw=black, fill=black, text=white, thick, minimum size=0.15pt, font=\tiny, inner sep=1.25pt] (current) {\textbf{1}}; points to the current syntax element in Python and R indicated by \tikz\draw node[circle, draw=black, fill=black, text=white, thick, minimum size=0.15pt, font=\tiny, inner sep=1.25pt] (current) {\textbf{2a}}; and \tikz\draw node[circle, draw=black, fill=black, text=white, thick, minimum size=0.15pt, font=\tiny, inner sep=1.25pt] (current) {\textbf{2b}};. Trevor looks over to the right at the explanation box:

\begin{center}
\includegraphics[width=\linewidth]{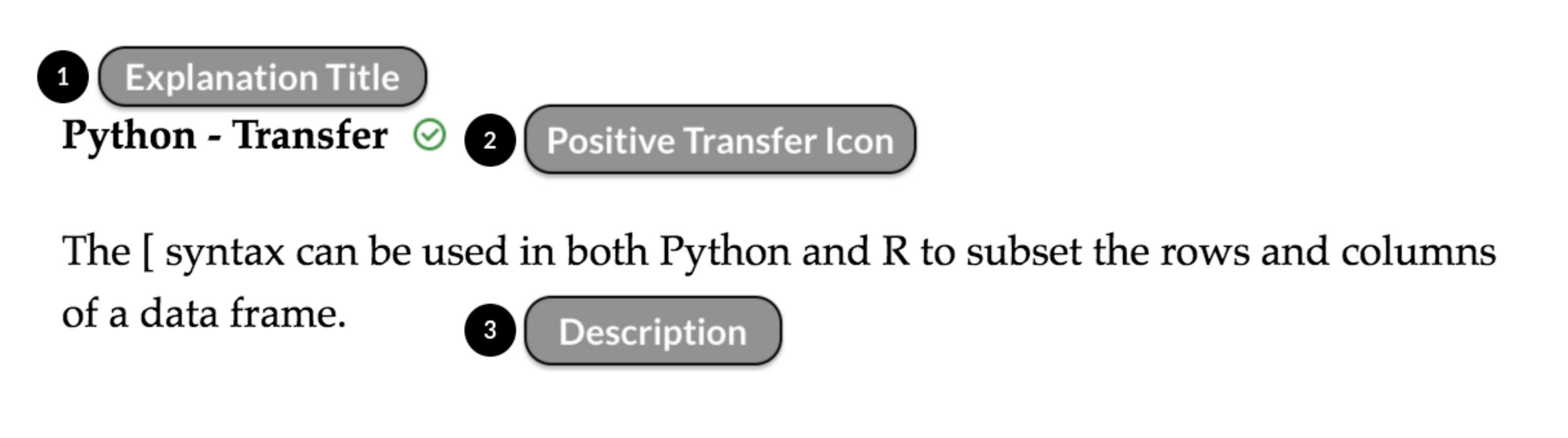}
\end{center}

% Below the snippets, Trevor looks at the other buttons and sees that there is also a way to traverse back an element with \tikz\draw node[circle, draw=black, fill=black, text=white, thick, minimum size=0.15pt, font=\tiny, inner sep=1.25pt] (current) {4};.

% To reduce the cognitive load on the programmer from having to read the whole line of code, the training tool has a stepper function to allow stepping through the relevant syntax elements piece by piece.

% The `Previous' button \tikz\draw node[circle, draw=black, fill=black, text=white, thick, minimum size=0.15pt, font=\tiny,inner sep=1.25pt] (current) {\textbf{4}}; highlights the previous element. The `Reset' button \tikz\draw node[circle, draw=black, fill=black, text=white, thick, minimum size=0.15pt,, font=\tiny,inner sep=1.25pt] (current) {\textbf{5}}; resets all the highlights and starts the lesson over. Finally, the `Done' button \tikz\draw node[circle, draw=black, fill=black, text=white, thick, minimum size=0.15pt, font=\tiny,inner sep=1.25pt] (current) {\textbf{6}}; finishes the lesson and moves on to the next one.
\subsubsection{Explanation Box}

Trevor sees \tikz\draw node[circle, draw=black, fill=black, text=white, text=white, thick, minimum size=0.15pt, font=\tiny,inner sep=1.25pt] (current) {\textbf{1}}; which refers to a Python `transfer' with \tikz\draw node[circle, draw=black, fill=black, text=white, text=white, thick, minimum size=0.15pt, font=\tiny,inner sep=1.25pt] (current) {\textbf{2}}; showing the transfer icon. He reads \tikz\draw node[circle, draw=black, fill=black, text=white, text=white, thick, minimum size=0.15pt, font=\tiny,inner sep=1.25pt] (current) {\textbf{3}}; and learns that the \texttt{[} operator can be used in R. \tool{} treats this syntax as a positive transfer since it can be reused. Trevor moves on to the next element:

\begin{center}
\includegraphics[width=\linewidth]{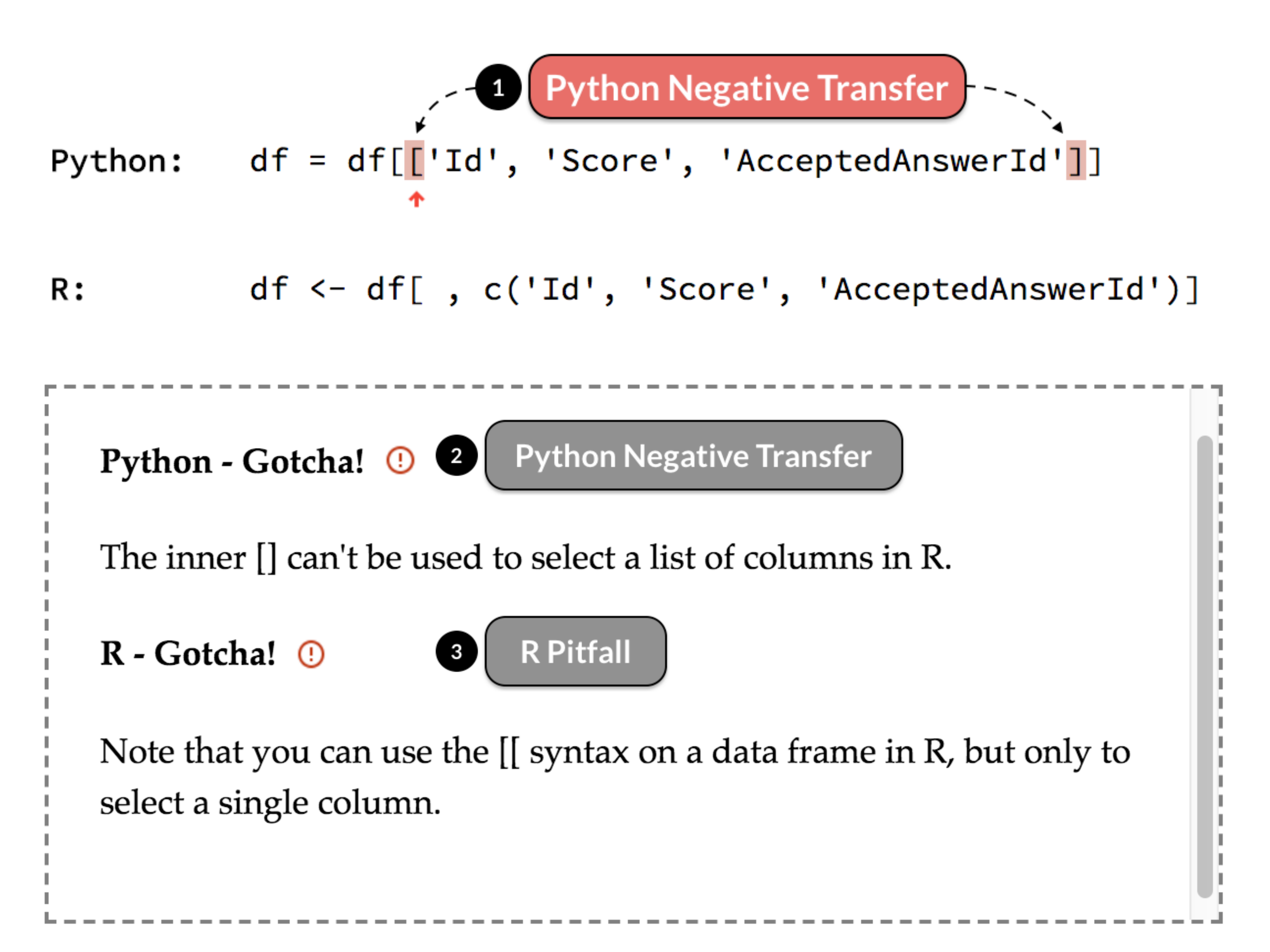}
\end{center}

% this paragraph then covers the gotcha with both the highlight and the explanation
Trevor looks at \tikz\draw node[circle, draw=black, fill=black, text=white, thick, minimum size=0.15pt, font=\tiny,inner sep=1.25pt] (current) {\textbf{1}}; which is a red highlight on the Python code. He reads \tikz\draw node[circle, draw=black, fill=black, text=white, thick, minimum size=0.15pt, font=\tiny,inner sep=1.25pt] (current) {\textbf{2}}; in the explanation box for clarification.

Trevor learns about a Python `gotcha': the \texttt{[[} syntax from Python can't be used in R. Trevor then reads \tikz\draw node[circle, draw=black, fill=black, text=white, thick, minimum size=0.15pt, font=\tiny,inner sep=1.25pt] (current) {\textbf{3}}; which explains an R `gotcha' about how the \texttt{[[} syntax is legal in R, but semantically different from the Python syntax as it only selects a single column. In this case, \tool{} warns him about a subtle difference, a negative transfer that could cause him issues in R. Trevor moves on to the next element and examines the elements that are highlighted blue:

\begin{center}
\includegraphics[width=\linewidth]{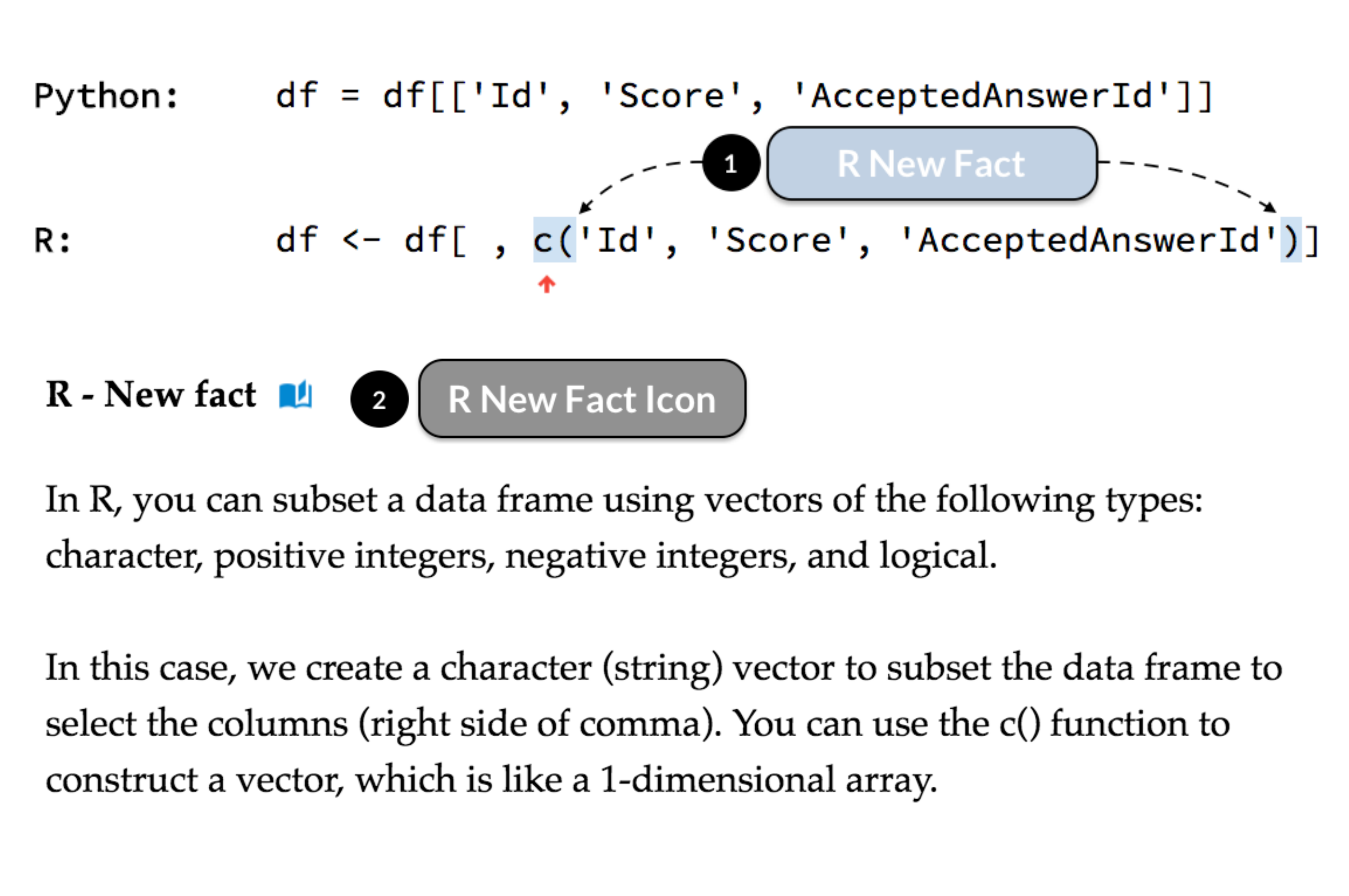}
\end{center}

Trevor looks at \tikz\draw node[circle, draw=black, fill=black, text=white, thick, minimum size=0.15pt, font=\tiny,inner sep=1.25pt] (current) {\textbf{1}}; then \tikz\draw node[circle, draw=black, fill=black, text=white, thick, minimum size=0.15pt, font=\tiny,inner sep=1.25pt] (current) {\textbf{2}}; and realizes he's looking at a `new fact' about R. \tool{} describes the \texttt{c()} function used to create a vector in R, which doesn't have a direct mapping to a Python syntax.

\subsubsection{Code Output Box}

Finally, Trevor steps through the code to the end, and the code output box now appear at the bottom which displays the state of the data frame:

\begin{center}
\includegraphics[width=\linewidth]{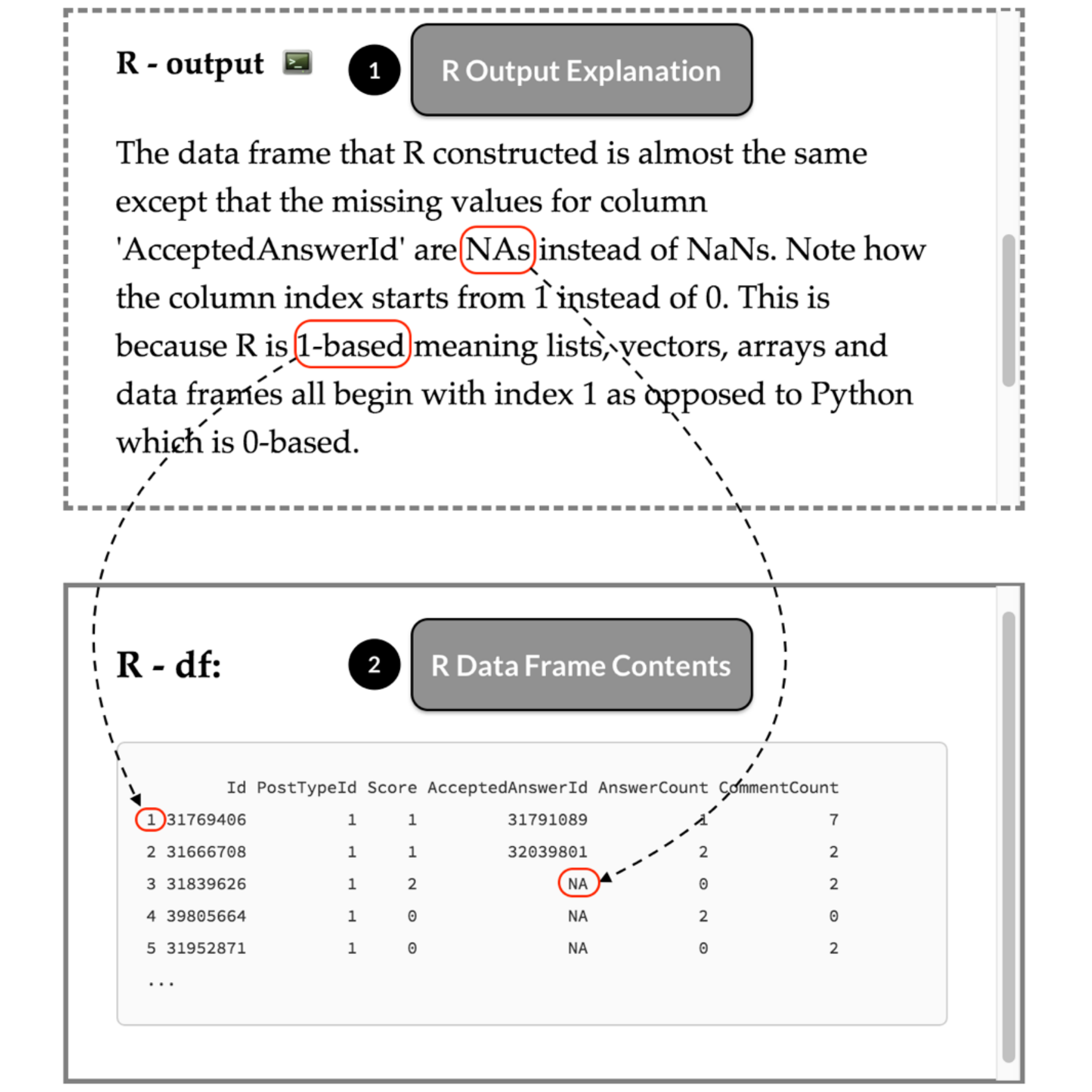}
\end{center}

Trevor reads \tikz\draw node[circle, draw=black, fill=black, text=white, thick, minimum size=0.15pt, font=\tiny,inner sep=1.25pt] (current) {\textbf{1}}; and inspects \tikz\draw node[circle, draw=black, fill=black, text=white, thick, minimum size=0.15pt, font=\tiny,inner sep=1.25pt] (current) {\textbf{2}}; to understand the contents of the data frame in R and how it differs from Python's data frame: 1) \texttt{NaNs} from Python are represented as \texttt{NAs} and 2) Row indices start from 1 as opposed to 0. \tool{} makes it clear that selecting columns of a data frame in R is similar to Python with some minor but important differences.

\section{Methodology}
\label{sec:methodology}

%\todo{Merge or delete this following paragraph.}
%In order to investigate the issue of transfer, we needed to compare two languages that illustrated different syntax and/or concepts for semantically equivalent code, resulting in transfer issues when making the transition from one to the other. Then, we needed to produce explanations of whether a syntax or concept from one language transfers positively or negatively to another to examine if that could help bridge the gap between the two. 

\subsection{Research Questions}
\label{sec:methodology:rqs}

% We investigated the following research questions through the traditional usability measures of effectiveness, usefulness, and user satisfaction~\cite{Dix2004}.

% feedback edit
We investigated three research questions using \tool{} to: 1) determine face validity of teaching a new language using an interactive tool 2) examine how programmers use \tool{} and 3) determine which affordances they found to being useful for learning a new language. 

% We investigated the following research questions through the traditional usability measures of effectiveness, usefulness, and user satisfaction~\cite{Dix2004}.

% \textbf{RQ1: Are programmers learning R through the tool?} We want to identify if training through learning transfer is an effective approach in the context of programming. A positive result of this research question validates the effectiveness of the tool.

\textbf{RQ1: Are programmers learning R through \tool{}?} To identify if training through learning transfer is an effective approach in the context of programming, this question is used to determine the face validity of \tool{}'s ability to teach R.

% By providing a mapping of their previous knowledge and warnings about negative transfers, the tool is aimed at helping programmers learn R efficiently but it is just as important that they actually learn the concepts covered in the lessons.

% \textbf{RQ2: How do programmers use the tool to support learning transfer?}
\textbf{RQ2: How do programmers use \tool{}?} Investigating how programmers use \tool{} can identify when it supports learning transfer, and whether the affordances in the tool align with the way programmers reason about the problem.

\textbf{RQ3: How satisfied are programmers with learning R when using the \tool{}?} We want to learn what features of \tool{} programmers felt were useful to them. If programmers are satisfied with the tool and find it useful, it is more likely to be used.

\subsection{Study Protocol}

\subsubsection{Participants}

\definecolor{sparkspikecolor}{named}{blue}

% We conducted a formative evaluation of a proof-of-concept training tool with 20 graduate students with an average age of 25 years (range of 22 to 43), four of whom were female.

% We used the programming languages Python and R for the study because both languages are used for data science programming tasks, yet have have subtle differences that are known to perplex novice R programmers with a background in Python~\cite{ohri2017python, burns:inferno, arggh}.

We recruited 20 participants from a graduate Computer Science course at our University, purposely sampling for participants with experience in Python, but not R. We chose to teach R for Python programmers because both languages are used for data science programming tasks, yet have have subtle differences that are known to perplex novice R programmers with a background in Python~\cite{burns:inferno, arggh, ohri2017python}.

Through an initial screening questionnaire, participants reported programming experience and demographics. Participants reported their experience with Python programming with a median of ``1-3 years'' (7), on a 4-point Likert-type item scale ranging from ``Less than 6 months'', ``1-3 years'', ``3-5 years'', and ``5 years or more'' (\begin{sparkline}{4}
\sparkspike .08 .5
\sparkspike .33 .35
\sparkspike .58 .05
\sparkspike .83 .1
\end{sparkline}). Participants reported a median of ``Less than 6 months'' (19) of experience with R programming (\begin{sparkline}{4}
\sparkspike .08 .95
\sparkspike .33 .05
\sparkspike .58 0.025
\sparkspike .83 0.025
% \sparkbottomlinex 0.58 0.83
\end{sparkline}), and reported a medium of ``1-3 years'' with data analysis activities (\begin{sparkline}{4}
\sparkspike .08 .75
\sparkspike .33 .15
\sparkspike .58 .05
\sparkspike .83 .025
\end{sparkline}). 16 participants reported their gender as male, and four as female; the average age of participants was 25 years ($sd$ = 5).

%Participants received extra credit for participating in the study. 

All participants conducted the experiment in a controlled lab environment on campus, within a 1-hour time block. The first author of the paper conducted the study.

% All participants had a university level programming experience and 10 of the participants had at least a year of experience in Python whereas only 1 participant had at a year or more experience in R. 

% There were 4 participants who had at least a year of experience with data analysis and 2 participants who had a year or more experience with Python's popular data analysis library called Pandas. 
% Previous work has shown that self estimation when compared to peers is a good measure of expertise compared to self estimation alone~\cite{feigenspan:experience} so we asked participants to estimate their expertise compared to their peers from a scale of 1 to 10, and the median response was 7.

%\subsubsection{Tasks}

\subsubsection{Onboarding}

Participants consented before participating in the study. They were presented with a general instructions screen which described the format of the study and familiarized them with the interface. The participants then completed a pre-test consisting of seven multiple choice or multiple answer questions, to assess prior knowledge on R programming constructs for tasks relating to indexing, slicing, and subsetting of data frames. The questions were drawn from our own expertise in the language and quizzes from an online text.\footnote{\url{http://adv-r.had.co.nz}, chapters ``Data Structures'' and ``Subsetting.''} The presentation of questions was randomized to mitigate ordering effects. We also asked participants to think-aloud during the study, and recorded these think-aloud remarks as memos.

% feedback addition
\subsubsection{Study Materials}

The authors designed four lessons on the topic of data frame manipulation, where each lesson consists of a one line code snippet in both languages and explanations associated with the relevant syntax elements. The authors also designed questions for the pre-test and post-test (see \cref{tbl:prepost}). Finally, the authors designed a user satisfaction survey of \tool{}. The study materials are available online.\footnote{\url{https://github.com/alt-code/Research/tree/master/TransferTutor}}
% \todo{link to repo}

\subsubsection{Tasks}

Participants completed the following lessons on R: 1) assignment and reading data, 2) selecting columns, 3) filtering, and 4) selecting rows and sorting. Participants stepped through each lesson as described in \cref{sec:thetool}. Within each lesson, participants interacted with 5--8 highlights and corresponding explanation boxes.

\subsubsection{Wrap-up} At the end of the study, participants completed a post-test containing the same questions as the pre-test. Participants completed a user satisfaction survey asking the participants for additional feedback on the tool. The survey asked them to rate statements about the usefulness of the tool using a 5-point Likert scale. These statements targeted different features of the tool such as whether or not highlighting syntax elements was useful for learning R. The survey also contained free-form questions for feedback regarding the tool such as the most positive and negative aspects, how they could benefit from using the tool and what features they would add to make it more useful. Finally, participants were given the opportunity to debrief for any general questions they may have had about the study.

\subsection{Analysis}

\begin{table}[t]
    \centering
    \begin{threeparttable}
    \caption{Pre-test and post-test questions\label{tbl:prepost}}
    \begin{tabularx}{\linewidth}{lXrr}
        \toprule
         \textbf{ID} & \textbf{Question Text} & \textbf{Tot.}\tnote{1} & \textbf{$\Delta$}\tnote{2}\\
         \midrule
         1 & Select all the valid ways of assigning a 1 to a variable `\texttt{x}' in R. & 0 & \cellcolor{green!25}18\\
         2 & Select all the valid vector types that can be used to subset a data frame. & 13 & \cellcolor{green!10}2\\
         3 & How would one check if `\texttt{x}' is the value \texttt{NA}? & 0 & \cellcolor{green!25}20\\
         4 & Given a data frame \texttt{df} with column indices 1, 2, and 3, which one of these will cause an error? & 10 & \cellcolor{green!10}3\\
         5 & Which one of these correctly selects the first row of a data frame \texttt{df}? & 0 & \cellcolor{green!25}20\\
         6 & Which one of these correctly subsets the first five rows and the first column of a data frame \texttt{df} and returns the result as a data frame? & 0 & \cellcolor{green!25}18\\
         7 & All of these statements correctly select the column `\texttt{c}' from a data frame \texttt{df} \emph{except} & 0 & \cellcolor{green!10}1\\
         \bottomrule
    \end{tabularx}
    \begin{tablenotes}
        % \item[1] Median score of pre-test.\\
        % \item[2] Median score increase between pre-test and post-test.
        \item[1] Total number of participants who answered correctly in pre-test.\\
        \item[2] Difference in the number of participants who answered correctly in pre-test and post-test.
    \end{tablenotes}
    \end{threeparttable}
\end{table}

\textbf{RQ1: Are programmers learning R through \tool{}?} We used differences in pre-test and post-test performance as a proxy measure for learning. We assigned equal weight to each question, with each question being marked as incorrect (0 points) or correct (1 point), allowing us to treat them as ordinal values. For the multiple answer questions, the participants received credit if they choose all the correct answers. A Wilcoxon signed-rank test between the participants' pre-test and post-test scores was computed to identify if the score differences were significant ($\alpha$ = 0.05).

% There was a pre test before the use of the tool and a post test after the final lesson. The questions tested the participants' knowledge on R using and were created using our own expertise in the language and some questions inspired from the `Data Structures' and `Subsetting' chapter quizzes from an online text called ``Advanced R"\footnote{\url{http://adv-r.had.co.nz}}.

% \textbf{RQ2: How do programmers use the tool to support learning transfer?}
\textbf{RQ2: How do programmers use \tool{}?}
All authors of the paper jointly conducted an open card sort---a qualitative technique for discovering structure from an unsorted list of statements~\cite{spencer2009card}. Our card sorting process consisted of two phases: \emph{preparation} and \emph{execution}. In the \emph{preparation} phase, we extracted the think-aloud and observational data from the written memos into individual cards, with each card containing a statement or participant observation. We labeled each of the cards as either being an indicator of positive transfer, negative transfer, or non-transfer. To do so, we used the following rubric to guide the labeling process:

\begin{enumerate}
    \item Statements should not be labeled if it includes verbatim or very close reading of the text provided by \tool{}.
    \item The statement can be labeled as positive if it demonstrates the participant learning a syntax or concept from Python that can be used in R.
    \item The statement can be labeled as negative if it demonstrates the participant learning a syntax or concept in R that is different from Python or breaks their expectation.
    \item The statement can be labeled as a non-transfer if it demonstrates the participant encountering a new fact in R for which there is no connection to Python.
\end{enumerate}

In the \emph{execution} phase, we sorted the cards into meaningful themes. The card sort is open because the themes were not pre-defined before the sort. The result of a card sort is not to a ground truth, but rather, one of many possible organizations that help synthesize and explain how programmers interact with tool.

\textbf{RQ3: How satisfied are programmers with learning R when using \tool{}?} We summarized the Likert responses for each of the statements in the user satisfaction survey using basic descriptive statistics. We also report on suggestions provided by participants in the free-form responses for questions, which include suggestions for future tool improvements.

\section{Results}

In this section we present the results of the study, organized by research question.

% \definecolor{Green}{HTML}{0f7f12}
% \definecolor{LightGreen}{HTML}{93ec94}
% \definecolor{Pink}{HTML}{fec0cb}
% \definecolor{Red}{HTML}{fc0d1b}

\newcommand{\likert}[4]{
\begin{minipage}[l]{\textwidth}
  \begin{tikzpicture}[xscale=0.0125, yscale=0.3]
      
    \node at (-100,0) {};
    \node at (100,0) {};

    \filldraw[color=Red] (#1, 0.0) rectangle (#2, 1.0);
    \filldraw[color=Pink] (#2, 0.0) rectangle (0, 1.0);
    \filldraw[color=LightGreen] (0, 0.0) rectangle (#3, 1.0);
    \filldraw[color=Green] (#3, 0.0) rectangle (#4, 1.0);

    \draw (0,0) -- (0, 1);
    
  \end{tikzpicture}
\end{minipage}    
}

\begin{table*}
\centering
\caption{Follow-up Survey Responses\label{tbl:usefulness}}
\begin{threeparttable}

\begin{tabularx}{\textwidth}{lrrrrrrX}
% \arrayrulecolor{white}
\toprule
& & \multicolumn{5}{c}{\textbf{Likert Resp. Counts$^1$}}\\
\cmidrule{3-7}

 & 
  \textbf{\% Agree} & 
  \textbf{SD} & 
  \textbf{D} & 
  \textbf{N} & 
  \textbf{A} & 
  \textbf{SA} & 
  \textbf{Distribution$^2$}\\  

\\ & & & & & & &

\begin{minipage}[l]{\textwidth}
  \begin{tikzpicture}[xscale=0.0125, yscale=0.3]    

    \node at (-100,0) {};
    \node at (100,0) {};

    \draw (-100,0) -- (100,0);
    \draw (-100,-0.25) -- (-100, 0.25);
    \draw (100,-0.25) -- (100, 0.25);    
    \draw (-50,-0.25) -- (-50, 0.25) node[above] {\tiny 50\%};
    \draw (50,-0.25) -- (50, 0.25) node[above] {\tiny 50\%};
    \draw (0,-0.25) -- (0, 0.25) node[above] {\tiny 0\%};    
    % % \node at (0, -0.28) {};
  \end{tikzpicture}
\end{minipage}\\

The highlighting feature was useful in learning about R. & 95\% & 0 & 0 & 1 & 5 & 14 &
\likert{0}{0}{25}{70}\\

Stepping through the syntax was useful in learning about R. & 79\% & 0 & 1 & 3 & 2 & 14 &
\likert{0}{-5}{10}{70}\\

The explanations that related R back to another language like Python was useful. & 89\% & 1 & 0 & 1 & 6 & 12 &
\likert{-5}{0}{30}{60}\\

The `new facts' in the information box helped me learn new syntax and concepts. & 95\% & 0 & 0 & 1 & 6 & 13 & 
\likert{0}{0}{30}{65}\\

The `gotchas' in the information box were helpful in learning about potential pitfalls. & 93\% & 0 & 2 & 0 & 6 & 12 & 
\likert{0}{-10}{30}{60}\\

The code output box helped me understand new syntax in R. & 79\% & 3 & 0 & 1 & 8 & 8 & 
\likert{-15}{0}{40}{40}\\

The code output box helped me understand new concepts in R. & 74\% & 2 & 0 & 3 & 7 & 8 & 
\likert{-10}{0}{35}{40}\\

\bottomrule
\end{tabularx}

\begin{tablenotes}
\item[1] Likert responses: Strongly Disagree (SD), Disagree (D), Neutral (N), Agree (A), Strongly Agree (SA). 
%Statements are worded such that strong agreement implies more challenging.
\item[2] Net stacked distribution removes the Neutral option and shows the skew between positive (more useful) and negative (less useful) responses. \\
\tikz \filldraw[color=Red] (0,0) rectangle (5pt,5pt); Strongly Disagree, \tikz \filldraw[color=Pink] (0,0) rectangle (5pt,5pt); Disagree, \tikz \filldraw[color=LightGreen] (0,0) rectangle (5pt,5pt); Agree; \tikz \filldraw[color=Green] (0,0) rectangle (5pt,5pt); Strongly Agree.
\end{tablenotes} 

\end{threeparttable}

\end{table*}

\subsection{RQ1: Are programmers learning R after using \tool{}?} 

% \todo{Nischal: check out the deal with "median"; I think it's total participants that got the question correct, then difference is between total who got it correct post vs pre tests.}

% RAW DATA
% Goodness-of-fit Shapiro-Wilk:
% Pre: W = 0.911742, P = 0.06
% Post: W = 0.792254, p = 0.0007
% Post test: 5.25 (u)
% Pre test: 2
% Mean difference: 3.25

% Wilcoxon signed rank test: S = 105, P < .0001.

All participants had a positive increase in overall score ($n$ = 20). The Wilcoxon signed rank test identified the post-test scores to be significantly higher than the pre-test scores ($S$ = 105, $p$ $<$ .0001), and these differences are presented in \cref{tbl:prepost}. Questions 1, 3, 5 and 6 provide strong support for learning transfer. In Question 2 and Question 4, most participants already supplied the correct answer with the pre-test: thus, there was a limited increase in learning transfer. The result of Question 7, however, was unexpected: no participants answered the pre-test question correctly, and there was essentially no learning transfer. We posit potential explanations for this in Limitations (\cref{sec:limitations}). Based on these results, using test performance has face validity in demonstrating \tool{}'s effectiveness in supporting learning transfer from Python to R.

% \subsection{RQ2: Did programmers use learning transfer?}
\subsection{RQ2: How do programmers use \tool{}?} The card sorting results of the observational and think-aloud memos are presented in this section, organized into four findings.

% \subsubsection{Participant spoke in terms of transfer}
\textbf{Evidence of using transfer:} We collected 398 utterances from our participants during their think-aloud during card sorting. All participants' think-aloud contained utterances related to learning transfer. 35.9\% of the total utterances related to transfer, revealing positive (18.9\%) and negative transfers (66.4\%). They also verbalized or showed behavior to indicate that they were encountering something that was new and didn't map to something they already knew (14\%). Other utterances not related to transfer involved verbatim reading of text or reflection on the task or tool.

% During the think-aloud, some participants spoke about how the elements might relate to one another in the two languages.
% Positive
% Participants identified several positive transfers from Python. One particular participant guessed that the range for selecting a column in the Python code was equivalent to the one in R without \tool{} explicitly mentioning this fact: \textit{``both are the same, 2 colon in Python means 3 in R.''} [P4]. Another participant correctly related Python's dot syntax to reference a data frame's column to R's use of dollar sign: \textit{``Oh looks like \$ sign is like the dot.''} [P17].

% evidence that programmers naturally make use of learning transfer, good sign for a tool that supports the strategy
Participants identified several positive transfers from Python, often without explicit guidance from \tool{}. P4 guessed that the range for selecting a column in the Python code was equivalent to the one in R without \tool{} explicitly mentioning this fact: \textit{``both are the same, 2 colon in Python means 3 in R.''} Another participant correctly related Python's dot notation to reference a data frame's column to R's use of dollar sign: \textit{``Oh looks like \$ sign is like the dot.''} [P17]. This is evidence that programmers are naturally using learning transfer and \tool{} helps support this strategy.

% some evidence of negative transfer from previous languages
Participants also encountered several negative transfers from either Python or their previous languages. P15 thought the dot in the \texttt{read.csv()} function signified a method call and verbalized that the \textit{``read has a csv function''} and later realized the mistake: \textit{``read is not an Object here which I thought it was!''} P5 expressed the same negative transfer, thinking that \textit{``R has a module called read.''}. This indicates a negative transfer from object-oriented languages where the dot notation is typically used for a method call.

% show evidence of non-transfer elements
Participants would also verbalize or show signs of behavior indicating that they have encountered a new fact, or a non-transfer, in R. This behavior occurred before progressing to the element with its associated explanation. P7 encountered the subsetting syntax in R and wondered, \textit{``Why is the left side of the comma blank?''} Another participant wondered about the meaning of a negative sign in front of R's \texttt{order} function by expressing they \textit{``don’t get why the minus sign is there.''} [P8].

% \todo{general formula}
% a) unpack the general observation statement, b) highlight some examples, reasons why this may be good or bad c) some future steps or alternatives to design.

\textbf{Tool highlighted facts participants may have misunderstood or missed:} The highlighting of the syntax elements and stepping through the code incrementally helped participants focus on the important parts of the code snippets. For additional feedback, one participant said \textit{``I was rarely confused by the descriptions, and the colorized highlighting helped me keep track of my thoughts and reference what exactly it was I was reading about with a specific example''} [P17]. P13 had a similar feedback remarking that the \textit{``highlighting was good since most people just try to summarize the whole code at once.''} However, a few participants found the stepper to progress the lesson too slowly. P17 read the entire line of code on the `Selecting rows and sorting' lesson and said that they \textit{``didn't understand drop=FALSE, hasn’t been mentioned''} before \tool{} had the opportunity to highlight it.

\textbf{Reluctance of accepting facts without execution or examples:} Participants were reluctant to accept certain facts without confirming for themselves through code execution, or without seeing additional examples. One participant was \textit{``not too sold on the explanation''} [P2] for why parentheses aren't required around conditions when subsetting data frames. Another participant expressed doubt and confusion when reading about an alternate \texttt{[} syntax that doesn't require specifying both rows and columns: \textit{``Ok but then it says you can use an alternate syntax without using the comma''} [P20]. Regarding the code output, one participant suggested that \textit{``it would’ve been more useful if I could change [the code] live and observe the output''} [P18]. There were a few participants who wanted more examples. For example, P17 was unclear on how to use the \texttt{[[} syntax in R and suggested that \textit{``maybe if there was a specific example here for the [[ that would help''}.

% Discussion-y:
% Participants were also interested in being able to compare the output changes associated with the code. This ties into the issue of a lack of code execution. Another issue that may have contributed to skepticism or lack of clarification was that some topics lacked examples of the concept being introduced. For example, one participant was unclear on how to use the \texttt{[[} syntax in R and suggested that \textit{``maybe if there was a specific example here for the [[ that would help"} [P17]. Future iterations of the tool could be redesigned to allow code execution and provide code examples in the text. This could help facilitate learning by allowing programmers to confirm the facts for themselves and reinforce what they have just learned.

% \subsubsection{Participants had difficulty with verbose and complex explanations}
\textbf{Information overload:} Although several participants reported that \tool{} is \textit{``interactive and easy to use''}  [P13], there were a few who thought that there was \textit{``information overload in the textual explanations''} [P1]. Some syntax elements had lengthy explanations and one participant felt that \textit{``sometimes too many new things were introduced at once''} [P18] and P5 expressed that \textit{``complex language is used''} to describe a syntax or concept in R. Participants also expressed that they wanted \textit{``more visual examples''} [P5].

\subsection{RQ3: How satisfied are programmers with learning R when using \tool{}?}
\label{sec:results:rq3}

Table \ref{tbl:usefulness} shows the distribution of responses for each statement from the user satisfaction survey, with each statement targeting a feature of \tool{}. Overall, participants indicated that features of \tool{} were useful in learning R. However, a few participants strongly disagreed about the usefulness of explanations relating back to R, and the output boxes. The free-form responses from participants offers additional insight into the Likert responses which will be discussed next.

The \emph{highlighting feature} had no negative ratings and all participants indicated that it was useful to them in some way. One participant thought that \textit{``the highlighting drew [their] attention''} [P2] while another commented that \textit{``it showed the differences visually and addressed almost all my queries''} [P1]. 

The \emph{stepper} received some neutral (3) ratings and one participant disagreed on its usefulness. Nevertheless, most participants did find the stepper useful and expressed that they \textit{``like how it focuses on things part by part''} [P20].

Participants generally found the \emph{explanations} relating R to Python was useful in learning R. One of the participants \textit{``liked the attempt to introduce R syntax based on Python syntax''} [P18] and P14 thought that \textit{``comparing it with Python makes it even more easy to understand R language''}. All participants thought this feature was useful except for one. This participant did not provide any feedback for why.

The `new facts' explanations also had no negative ratings and was useful to all participants. Although participants didn't speak explicitly about the feature, P8 expressed that there was \textit{``detailed explanation for each element''} and P16 said that \textit{``Every aspect of the syntax changes has been explained very well''}. Most participants also found `gotchas' to be useful. P7 for example said that \textit{``Gotchas! were interesting to learn and to avoid errors while coding.''}

For the \emph{explanation box}, some participants suggested that this affordance would need to \textit{``reduce the need for scrolling and (sadly) reading''} [P2]. Still other participants wanted deeper explanations for some concepts, perhaps with \textit{``links to more detailed explanations''} [P12]. For the \emph{output boxes}, participants who disagreed with its usefulness suggested that the output boxes would be more useful if the output code be dynamically adjusted by changing the code [P6, P9, P12], and P17 suggested that the output boxes were \textit{``a little difficult to read''} because of the small font.

\section{Limitations}
\label{sec:limitations}

% The selection of code snippets and programming languages, along with the qualitative research approaches used in our think-aloud study introduce trade-offs in the design and reporting of our study.

\subsection{Construct Validity} 
% We used pre-test and post-test questions as a proxy to assess the participants' understanding of R concepts as covered by the tool. Because of time constraints in the study, we can only ask a limited number of questions. Consequently, these questions are only approximations of participants' understanding. For instance, Question 7 illustrates several reasons why questions may be problematic for programmers. First, the question may be confusingly-worded, because of the use of \emph{except} in the question statement. Second, the response may be correct, but incomplete---due to our scoring strategy, responses must be completely correct to receive credit. Third, questions are only approximations of the participants' understanding.

% camera-ready edit
We used pre-test and post-test questions as a proxy to assess the participants' understanding of R concepts as covered by \tool{}. Because of time constraints in the study, we could only ask a limited number of questions. Consequently, these questions are only approximations of participants' understanding. For instance, Question 7 illustrates several reasons why questions may be problematic for programmers. First, the question may be confusingly-worded, because of the use of \emph{except} in the question statement. Second, the response may be correct, but incomplete---due to our scoring strategy, responses must be completely correct to receive credit. Third, questions are only approximations of the participants' understanding. A comparative study is necessary to properly measure learning from using \tool{} to other traditional methods of learning languages by measuring performance on programming tasks.

\subsection{Internal Validity} Participants in the study overwhelmingly found the features of \tool{} to be positive (\cref{sec:results:rq3}). It's possible, however, that this positivity is artificially high due to social desirability bias~\cite{Dell2012}---a phenomenon in which participants tend to respond more positively in the presence of the experimenters than they would otherwise. Given the novelty of \tool{}, it is likely that they assumed that the investigator was also the developer of the tool. Thus, we should be conservative about how we interpret user satisfaction with \tool{} and its features.

A second threat to internal validity is that we expected \tool{} to be used by experts in Python, and novices in R. Although all of our participants have limited knowledge with R, very few participants were also experts with Python or the Pandas library (\cref{sec:methodology}). On one hand, this could suggest that learning transfer would be even more effective with expert Python/Pandas participants. On the other hand, this could also suggest that there is a confounding factor that explains the increase in learning that is not directly due to the tool. For instance, it may be that explanations in general are useful to participants, whether or not they are phrased in terms of transfer~\cite{Kulesza2013,Bunt2012}.

\subsection{External Validity} We recruited graduate students with varying knowledge of Python and R, so the results of the study may not generalize to other populations, such as industry experts. The choice of Python and R, despite some notable differences, are both primarily intended to be used as scripting languages. How effective language transfer can be when language differences are more drastic is still an open question; for example, consider if we had instead used R and Rust---languages with very different memory models and programming idioms.

\section{Design Implications}
\label{sec:implications}

This section presents the design implications of the results and future applications for learning transfer.

\subsection{Affordances for supporting learning transfer}

% First, something about how tool helped programmers focus and catch negative transfer (pros/cons)

Stepping through each line incrementally with corresponding highlighting updates allows programmers to focus on the relevant syntax elements for source code. This helps novice programmers pinpoint misconceptions that could be easily overlooked otherwise, but prevents more advanced programmers from easily skipping explanations from \tool{}. Despite the usefulness of always-on visualizations in nice environments~\cite{kang2017omnicode,Hoffswell:2018:ACS:3173574.3174106}, an alternative implementation approach to always-on may be to interactively allow the programmer to activate explanations on-demand.

We found that live code execution is an an important factor for programmers as they can test new syntax rules or confirm a concept. We envision future iterations of \tool{} that could allow code execution and adapt explanations in the context of the programmers' custom code.

% Finally, it's important to reduce information overload and balance the volume of explanation against the amount of code to be explained. One solution is to externalize additional explanation to documentation outside of the tool, such as web resources. Breaking up lessons into smaller segments could also reduce the amount of reading required.

% camera-ready edit
% making this paragraph just as another point
Reducing the amount of text and allowing live code execution were two improvements suggested by the participants. This suggests that  \tool{} needs to reduce information overload and balance the volume of explanation against the amount of code to be explained. One solution is to externalize additional explanation to documentation outside of \tool{}, such as web resources. Breaking up lessons into smaller segments could also reduce the amount of reading required.

% addressing authoring issues mentioned by reviewers
% Finally, one of the limitations of \tool{} is the manual work required to author the lessons, which could be mitigated by automating the analysis of code snippets and generating lessons. We plan to explore ways to automatically analyze code snippets in two programming languages and provide similarities and differences between syntax elements to assist the lesson designer. An interface to create new lessons would also be helpful for instructors.

\subsection{Expert learning can benefit from learning transfer}

% The \emph{expertise reversal effect} suggests that instructional techniques that are effective for novices can have negative consequences for experience learners~\cite{kalyuga:expertise}. In contrast to explanations for novices, we intentionally mitigated the expertise reversal effect by presenting explanations in terms of language transfer---in the context of a language that the programmer is already an expert at. We think that tools such as ours can activate as a type of intervention design: like training wheels, programmers new to the language can use our tool to familiarize themselves with the language. As they become experts, they would reduce and eventually eliminate use of the tool.

% camera ready edit
% addressing: 
% - (R1) "we need a better experimental setup comparing this approach to a programming education approach for 100% novices"
% - R3 suggested improvement: "there should be more discussion about which aspects of programming knowledge are easy and hard to transfer."
To prevent negative consequences for experienced learners, we intentionally mitigated the expertise reversal effect~\cite{kalyuga:expertise} by presenting explanations in terms of language transfer---in the context of a language that the programmer is already an expert at. Participants in our study tried to guess positive transfers on their own, which could lead to negative transfers from their previous languages. This cognitive strategy is better supported by a tool like \tool{} as it guides programmers on the correct positive transfers and warns them about potential negative transfers. We think that tools such as ours serves as a type of intervention design: like training wheels, programmers new to the language can use our tool to familiarize themselves with the language. As they become experts, they would reduce and eventually eliminate use of \tool{}.

% A controlled study is required to properly measure learning gains when using the transfer approach versus traditional methods. It is also unclear what types of programming knowledge are easy and hard to transfer. Further investigations are required to identify the types of transfer issues programmers face and the relative difficulty of each type.

\subsection{Learning transfer within programming languages}
% https://rviews.rstudio.com/2017/06/08/what-is-the-tidyverse/

Our study explored learning transfer between programming languages, but learning transfer issues can be found within programming languages as well, due to different programming idioms within the same language. For example, in the R community, a collection of packages called \texttt{tidyverse} encourage an opinionated programming style that focuses on consistency and readability, through the use of a \emph{fluent} design pattern. In contrast to `base' R---which is usually structured as a sequence of data transformation instructions on data frames---the fluent pattern uses `verbs' that pipe together to modify data frames.

\subsection{Applications of learning transfer beyond tutorials}

Learning transfer could be applied in other contexts, such as within code review tools, and within integrated development environments such as Eclipse and Visual Studio. For example, consider a scenario in which a software engineer needs to translate code from one programming language to another: this activity is an instance in which learning transfer is required. Tools could assist programmers by providing explanations in terms of their expert language through existing affordances in development environments. Learning transfer tools can be beneficial even when the language conversion is automatic. For example, SMOP (Small Matlab and Octave to Python compiler) is one example of a transpiler---the system takes in Matlab/Octave code and outputs Python code.\footnote{\url{https://github.com/victorlei/smop}} The generated code could embed explanations of the translation that took place so that programmers can better understand why the translation occurred the way that it did.

Another potential avenue for supporting learning transfer with tools can be found in the domain of documentation generation for programming languages. Since static documentation can't support all types of readers, authors make deliberate design choices to focus their documentation for certain audiences. For example, the canonical Rust book\footnote{\url{https://doc.rust-lang.org/book/second-edition/}} makes the assumption that programmers new to Rust have experience with some other language---though it tries not to assume any particular one. Automatically generating documentation for programmers tailored for prior expertise in a different language might be an interesting application for language transfer.

\section{Related Work}

%\todo{Titus: talk about ~\cite{Campbell1992} ?}

%\subsection{Programming Language Transfer Studies}
%There is a large body of work that has focused on transfer between text editors, based on the common elements theory of transfer, which states that knowledge of one context will transfer to another context only if the two share elements~\cite{polson:quantitative, polson:transfer, polson:test, singley:keystroke}. 
There are many studies on transfer between tools~\cite{polson:quantitative, polson:transfer, polson:test, singley:keystroke} but fewer studies examining transfer in programming. Transfer of declarative knowledge between programming languages has been studied by Harvey and Anderson~\cite{harvey:transfer}, which showed strong effects of transfer between Lisp and Prolog. Scholtz and Wiedenbeck~\cite{Scholtz:subsequent} conducted a think-aloud protocol where programmers who were experienced in Pascal or C tried implementing code in Icon. They demonstrated that programmers could suffer from negative transfer of programming languages. Wu and Anderson conducted a similar study on problem-solving transfer, where programmers who had experience in Lisp, Pascal and Prolog wrote solutions to programming problems~\cite{quanfeng}. The authors found positive transfer between the languages which could improve programmer productivity. Bower~\cite{Bower:2011:CEC:1999747.1999809} used a new teaching approach called Continual And Explicit Comparison (CAEC) to teach Java to students who have knowledge of C++. They found that students benefited from the continual comparison of C++ concepts to Java. However, none of these studies investigated tool support.

Fix and Wiedenbeck~\cite{Fix} developed and evaluated a tool called ADAPT that teaches Ada to programmers who know Pascal and C. Their tool helps programmers avoid high level plans with negative transfer from Pascal and C, but is targeted at the planning level. Our tool teaches programmers about negative transfers from Python, emphasizing both syntax and semantic issues by highlighting differences between the syntax elements in the code snippets of the two languages. \tool{} also covers pitfalls in R that doesn't relate to Python.  %Our tool is also not designed for writing R code.

% The tool was designed to support implementation planning~\cite{Scholtz:subsequent} in Ada to solve problems. Programmers could select high level plans with a graphical menu and once the lowest level plan was selected they could type out Ada code.  Unfortunately, the programmers in the study still struggled with Ada syntax because of negative transfers. In response to an incorrect Ada code, ADAPT presents an error message in a dialog window teaching the programmers about the correct code to use instead. Like ADAPT, our tool also helps programmers avoid negative transfers from their previous languages. However, 

% Nischal: this can be good to point out as an inspiration for the stepping function!
We leverage existing techniques used in two interactive learning tools for programming, namely Python Tutor~\cite{guo:tutor} and Tutorons~\cite{head:tutorons}. Python Tutor is an interactive tool for computer science education which allows the visualization and execution of Python code. We borrowed the idea of Python Tutor's ability to step through the code and pointing to the current line the program is executing to help the programmer stay focused. Head et al. designed a new technique of generating explanations or \emph{Tutorons} that helps programmers learn about code snippets on the web browser by providing pop-ups with brief explanations of user highlighted code~\cite{head:tutorons}. Although our tool does not automatically generate explanations for highlighted code, it uses the idea of providing details about syntax elements as the programmer steps through the syntax elements which are already highlighted for them.

\section{Conclusion}

In this paper, we evaluated the effectiveness of using learning transfer through a training tool for expert Python developers who are new to R. We found that participants were able to learn basic concepts in R and they found \tool{} to be useful in learning R across a number of affordances. Observations made in the think-aloud study revealed that \tool{} highlighted facts that were easy to miss or misunderstand and participants were reluctant to accept certain facts without code execution. The results of this study suggest opportunities for incorporating learning transfer feedback in programming environments.

\section*{Acknowledgements} This material is based in part upon work supported by the National Science Foundation under Grant Nos. 1559593 and 1755762.

% \newpage
% \raggedright

\bibliographystyle{IEEEtran}
\IEEEtriggeratref{20}
\bibliography{references}

\end{document}